\documentclass[final]{raa06}           
\usepackage{graphicx,times}             
\usepackage{natbib} 
\usepackage{amssymb,amsmath}
\bibpunct{(}{)}{;}{a}{}{,}
\usepackage[colorlinks=true, citecolor=blue]{hyperref}%

\usepackage{graphicx,kantlipsum,setspace}
\usepackage{caption}
\usepackage{orcidlink} 
\usepackage[T1]{fontenc}
\usepackage{ae,aecompl}

\usepackage{graphics,epsf}
\usepackage{amsmath}                
\usepackage{amsfonts}               
\usepackage{amssymb}                
\usepackage{epsfig}                 
\usepackage{appendix}
\usepackage{graphicx}
\usepackage{float}
\usepackage{color}
\usepackage{multirow}
\usepackage{colortbl}
\usepackage[para,online,flushleft]{threeparttable}
\usepackage{xcolor}

\hypersetup{citecolor=blue, 
            linkcolor=red, 
            menucolor=blue, 
            urlcolor=blue}  

\newcommand{\cm}{{~\rm cm}}

\newcommand{\km}{{~\rm km}}
\newcommand{\s}{{~\rm s}}

\newcommand{\g}{{~\rm g}}

\newcommand{\K}{{~\rm K}}
\newcommand{\erg}{{~\rm erg}}
\newcommand{\yr}{{~\rm yr}}

\begin{document}

\title{Building three-dimensional giant stellar models for common envelope simulations}



   \volnopage{Vol.0 (20xx) No.0, 000--000}      
   \setcounter{page}{1}          


   \author{Ron Schreier, Shlomi Hillel, Noam Soker\,\orcidlink{0000-0003-0375-8987} 
    }  

   \institute{Department of Physics, Technion - Israel Institute of Technology, Haifa, 3200003, Israel;   {\it    shlomi.hillel@gmail.com; ronsr@technion.ac.il,; soker@technion.ac.il}\\
\vs\no
   {\small Received~~2026 month day; accepted~~2026~~month day}}

\abstract{
We build a three-dimensional (3D) red supergiant (RSG) stellar model by transporting a 1D stellar model to a 3D numerical grid, mimicking core nuclear power by depositing energy into an inner shell, and mimicking stellar emission by cooling grid cells with densities below the photospheric density. The goal is to use such models for common envelope evolution (CEE) and grazing envelope evolution simulations. We do not relax the model; rather, we let it perform its natural pulsation. We find that when we mimic photospheric emission by cooling low-density grid cells, the oscillations decay slowly on a timescale much longer than without photospheric cooling. When we mimic both nuclear energy production, by depositing the stellar luminosity in an inner shell above the inert core of the stellar model, and photospheric cooling, the oscillations do not decay; instead, their amplitude slowly increases with time. The main pulsational period is about 1 year, comparable to the stellar dynamical time, suggesting a fundamental radial pulsation mode. The non-spherical structure of the stellar model and rapid low-amplitude temporal variations in the average stellar radius tentatively hint at the presence of non-radial oscillation modes on top of the fundamental radial mode. We also obtain convection, as RSG stars should have. We conclude that energy deposition in the inner envelope at the stellar luminosity and the cooling of the outer low-density numerical cells are a way of preparing a pulsating giant star to simulate CEE and grazing-envelope evolution. This procedure requires no relaxation of the stellar model. 
}

\maketitle

\section{Introduction}
\label{sec:Introduction}

The large dynamical range in the common envelope evolution (CEE) process and its three-dimensional (3D) nature make CEE numerical simulations prohibitively complicated (e.g., a review by \citealt{RoepkeDeMarco2023}). Consider a CEE of a small star, the secondary, which might be a black hole, a neutron star, a white dwarf, or a main-sequence star that spirals inside the bloated envelope of an evolved star, the primary, i.e., a red giant branch (RGB), an asymptotic giant branch (AGB), or a red supergiant (RSG) star. The numerical grid should resolve the large common envelope, which can expand to dimensions much larger than those of the initial giant star. It should also resolve the small star to a resolution smaller than its radius. In between, the numerical code should resolve convection motion in the envelope, and the accretion process onto the companion that starts at the accretion radius and continues to the surface of the secondary star. The most demanding requirement for the numerical code is to resolve the accretion process onto the secondary star, which may occur via an accretion disk that launches jets. No existing numerical code fulfills this demand. 
 
The claim for major roles of jets in the CEE comes mainly from planetary nebulae. Over a hundred post-AGB nebulae and planetary nebulae have a close central binary system, namely, a post-CEE binary system (e.g., \citealt{Miszalski2019ic, Oroszetal2019, Jones2020Galax, Jones2025}). Some may be post-grazing envelope-evolution binary systems. 
The large number of planetary nebulae that exhibit axisymmetrical (one symmetry axis) or point-symmetric (multipolar; i.e., two or more symmetry axes) morphologies (e.g., \citealt{Balick1987, Chuetal1987, Schwarzetal1992, CorradiSchwarz1995, SahaiTrauger1998, Sahaietal2007, Sahaietal2011, Parkeretal2016, Parker2022}, for papers and catalogs with large collections of planetary nebula images) testify to the major role of jets in the CEE, or shortly before the CEE or shortly after it   
(e.g., \citealt{Morris1987, Soker1990AJ, GarciaSegura1997, SahaiTrauger1998, GarciaSeguraLopez2000, GarciaSeguraetal2005, BlackmanLucchini2014, Tocknelletal2014, GarciaSeguraetal2021, GarciaSeguraetal2022, AkashiSoker2018, EstrellaTrujilloetal2019, Tafoyaetal2019, Balicketal2020, RechyGarciaetal2020, Clairmontetal2022, Danehkar2022, MoragaBaezetal2023, Derlopaetal2024, Mirandaetal2024, Sahaietal2024}; \citealt{Baanetal2021} discussed an alternative shaping process). \cite{Soker2025Robust} argued that jets, which the secondary star launches as it accretes mass from the common envelope, are the most robust observation of the CEE, besides the presence of the close binary system that shows the system has gone through a CEE. The presence of an inflated main-sequence companion to the central star in some planetary nebulae indicates that the companion has accreted mass during the CEE (e.g., \citealt{Jonesetal2015}).

Another indication of the role of jets comes from studies that conclude that the orbital energy released by the binary system during the CEE alone was insufficient to eject the common envelope (e.g., \citealt{Grichener2023, LiZhenweietal2026CE}). Jets that carry accretion energy from the secondary star can facilitate envelope removal. We accept the view that the standard CEE model should include jets. Namely, the standard CEE should not be the traditional one (e.g., \citealt{Paczynski1976, Webbink1984}) in which the only energy source is orbital energy.    

Although several CEE simulations in the last decade do include jets (e.g., \citealt{MorenoMendezetal2017, ShiberSoker2018, LopezCamaraetal2019, Shiberetal2019, LopezCamaraetal2020MN, Hilleletal2022, Hilleletal2023, LopezCamaraetal2022, Zouetal2022, Soker2022Rev, Schreieretal2023, Schreieretal2025, Gurjareta2024eas, ShiberIaconi2024}), the majority do not include jets  (e.g., \citealt{Staffetal2016MN, Kuruwitaetal2016, Ohlmannetal2016a,  Iaconietal2017b, Chamandyetal2019, LawSmithetal2020, GlanzPerets2021a, GlanzPerets2021b, GonzalezBolivaretal2022, GonzalezBolivaretal2024, Lauetal2022a, Lauetal2022b,  BermudezBustamanteetal2024, Chamandyetal2024, GagnierPejcha2024,  Landrietal2024, RosselliCalderon2024, Vetteretal2024, Bhattacharyyaetal2025, Vetteretal2025, Gagnieretal2026}).  
Due to numerical complexity, jetted simulations of the CEE, namely those in which the companion launches the jets, omit other processes, such as accretion by the companion or its gravity, or they cover only a short evolutionary time. 

Modeling the envelope density correctly is important, as it determines the mass accretion rate onto the secondary star. This is particularly important when the secondary enters the envelope because these giant stars pulsate. In this study, we build an RSG stellar model that mimics the star's luminosity from nuclear burning in the core and energy loss from its photosphere (Section \ref{sec:Methods}). We will find that we can reproduce stellar pulsations (Section \ref{sec:Pulsation}). We compare one regular-resolution simulation with a high-resolution simulation, but limited in time, in Section \ref{sec:Convergence}.  In Section \ref{sec:Summary}, we summarize by concluding that there is no need to relax the giant envelope for CEE simulations; it is possible to let the stellar model pulsate. 

\section{Numerical methods and assumptions}
\label{sec:Methods}

\subsection{New numerical ingredients}
\label{subsec:Ingredients}
We transport a 1D stellar model to a 3D grid. Many studies that transfer a 1D model to a 3D grid relax the stellar model (damping velocities) to start the CEE simulation with an envelope velocity of zero (e.g., \citealt{RickerTaam2008, 
Passyetal2012, Chamandyetal2018, Ohlmannetal2017,PrustChang2019, GlanzPerets2021a, GlanzPerets2021b}; see review by \citealt{RoepkeDeMarco2023}; more in Section \ref{subsec:Relaxed}). 
We, instead, took the approach of not relaxing the model, but rather letting it pulsate \citep{Hilleletal2023, Schreieretal2025}. However, these pulsations decay after about two periods, and the star expands.  

In this study, we add two new ingredients to our 3D stellar model. 
(1) We mimic the radiation from the stellar surface by cooling the gas with densities below that of the initial photosphere density. 
(2) We mimic the nuclear energy generation in the core by depositing energy in an inner shell at the luminosity of the 1D stellar model. 

\subsection{Numerical code and stellar model}
\label{subsec:Code}

Our numerical setup is similar to our previous papers, e.g., \citep{Hilleletal2023, Schreieretal2025}. 
There are three new features: the energy deposition and cooling that we described in Section \ref{subsec:Ingredients}, and the imposed spherical outer boundary (see below).

We built an RSG stellar model from a zero-age-main-sequence star of metallicity $Z=0.02$ and mass $M_{\rm 1,ZAMS}=15 M_\odot$, using the 1D stellar evolution code \texttt{MESA} \citep{Paxtonetal2011, Paxtonetal2013, Paxtonetal2015, Paxtonetal2018, Paxtonetal2019}. We transport the 1D RSG stellar model, with a mass of $M_1=12.5 M_\odot$ and a radius of $R_{\rm RSG}=881\,R_{\odot} = 6.1 \times 10^{13} \cm$, into our 3D numerical grid of the hydrodynamical numerical code {\sc flash} \citep{Fryxelletal2000}. To save computational time, we keep an inner inert sphere with radius $R_{\rm inert} = 0.2R_{\rm RSG} = 176\,R_{\odot}$ and mass $5.64 M_\odot$; we do not change its structure. 
This radius is chosen considering two factors. The first is numerical. Taking a smaller radius for the inert core would require following inner envelope zones where the dynamical time is shorter; hence, time steps might need to be shorter, and the computation too expensive. The second one is of interest to us. We are interested in simulating jets at the onset of the CEE and its early phases, and in simulating the grazing envelope evolution. We are not aiming at a companion that spirals in deep into the envelope. Studies interested in later phases of the CEE will have to use a smaller inert core.

We deposit the energy that mimics the stellar nuclear burning as close as we can to the center. Since we exclude the inner inert sphere of radius $176R_\odot$, we deposit the energy just above this radius, within the shell  $14\times10^{12} \cm = 201 R_\odot < r < 230 R_\odot = 16 \times 10^{12} \cm$. This shell is two cells wide.  

We include the gravity of the RSG star as its value when we start the 3D simulations, including the gravity of the inert inner sphere. We use a Cartesian grid and assume the gas is an ideal gas with an adiabatic index of $\gamma = 5/3$, including radiation pressure. 
The center of the RSG is at the origin. The cell size is $\Delta _{\rm c}=L_{\rm G}/256 = 9.766 \times 10^{11} \cm$, for the edge of the cubical grid of $L_{\rm G} = 250 \times 10^{12} \cm$. 

In this study, we prevent the influence of the cubical boundary of the grid by imposing a spherical numerical boundary at $R_{\rm w}=100 \times 10^{12} \cm = 1437 R_\odot= 1.63 R_{\ast,0}$, where $R_{\ast,0}$ is the initial radius of the 1D stellar model. At this radius, we expect a well-defined stellar wind to exist, which we do not include in this study. From this radius outward, we set all velocities to a radial outflow $v=0.1 \km \s^{-1}$. In future studies, we will explore different outer boundary conditions and find the best one to mimic the stellar wind.   

We do not include radiative transport because it is too computationally intensive. To mimic the star's radiation, we remove energy from the low-density regions. We set a `numerical photospheric density,' $\rho_{\rm ph}$, and at each time step we check all numerical cells with a density of $\rho < \rho_{\rm ph}$ and reduce their temperature to a `numerical chromospheric temperature' $T_{\rm ch}$. In the present study, we set $\rho_{\rm ph}=2\times 10^{-9} \g \cm^{-3}$, based on the 1D photospheric density of $2.1 \times 10^{-9} \g \cm^{-3}$, and $T_{\rm ch}= 1000 \K$. 

To mimic the nuclear power of the stellar core, we inject energy at a power equal to the 1D stellar luminosity $L= 2.65 \times 10^{38} \erg \ s^{-1} \simeq 7 \times 10^4 L_\odot$, into a shell bounded by $14 \times 10^{12} \cm = 201 R_\odot < r  < 16 \times 10^{12} \cm = 230 R_\odot $, which is above and close to the inert core.  
We will present two simulations: one with and one without mimicking nuclear energy. 

We do not include recombination energy of hydrogen or helium in the envelope, nor mass loss, which might be significant in RSG (e.g., \citealt{Senguptaetal2026}). We expect CEE-induced energy and mass loss to be much more important than recombination energy and pulsation-driven mass loss. Our goal is to prepare giant stars for CEE simulations.    

\subsection{Comparison with relaxed stellar model}
\label{subsec:Relaxed}

In their review, \cite{RoepkeDeMarco2023} summarized the construction of relaxed giant stellar models for CEE simulations in their section 5.3. Essential to this approach is relaxing the 3D stellar model imported from a 1D high-resolution stellar model (e.g., \citealt{RickerTaam2008, Passyetal2012, Ohlmannetal2016a, Ohlmannetal2017, Chamandyetal2018, PrustChang2019, Reichardtetal2020}).

\cite{Ohlmannetal2017} listed five quality criteria for the representation of primary
stars in 3D hydrodynamic CEE simulations (adopted by \citealt{RoepkeDeMarco2023} in their review):
(1) The deviations from the initial pressure and density profiles should be small.
(2) Both sides of the hydrostatic equilibrium equation should compensate each other.
(3) For convectively stable profiles, the Mach numbers in the model should be small compared to the Mach numbers expected in the problem. 
(4) For convectively unstable models, a stationary state should be reached.
(5) To exclude the presence of pulsations, the potential energy of the system should be constant.

We import the 1D stellar model to the 3D grid; hence, we also require that the pressure and density profiles be similar to the initial ones (point 1). Point (3) is not relevant to us, as we deal only with pulsating giant stars that have convective envelopes, and we simulate only the envelope. We differ in points 2, 4, and 5. We do not require that the hydrostatic equation holds (point 2), we do not require a stationary state (point 4), and we do not exclude pulsation (point 5). On the contrary, we aim to have large pulsations. However, as we will show later, the pulsations in radii of mass elements and energy are around more or less constant values. So the average values largely reach an equilibrium. 

Some studies that use the relaxed stellar approach (e.g., \citealt{Iaconietal2018}) prevent any motion of the 3D stellar model by setting the pressure in the grid cells outside the star to have the same pressure as the stellar surface. We, on the other hand, reduce the temperature (hence the pressure) of the outer cells to mimic the stellar radiation. 

We do not claim that our approach to building a giant stellar model for CEE simulations is better than the more traditionally relaxed stellar model approach. We only claim that the pulsations and convection we obtained, which the relaxed stellar model does not have, better reflect the behavior of giant stars, particularly RSG stars. For example, the progenitor of SN~2023ixf had large-amplitude pulsations before explosion (e.g., \citealt{Kilpatricketal2023, Soraisametal2023}). Such pulsations allow better inclusion of the onset of the CEE and convection-driven energy transport, which is crucial for accurate CEE simulations (e.g., \citealt{WilsonNordhaus2020, WilsonNordhaus2022, Grondinetal2026}).

\section{Stellar pulsation}
\label{sec:Pulsation}

The key approach is not to relax the model but rather to let the stellar model pulsate.
The new elements we introduce (Section \ref{subsec:Ingredients}) mimic two processes. (1) By reducing the temperature of low-density cells, i.e., lower than about the photospheric density, we mimic the photospheric cooling by regular stellar radiation. (2) By depositing energy in an inner shell above the inert core, we mimic the stellar nuclear energy production in the core. 
We will present results for two basic cases: with both ingredients included and with only photospheric cooling included. In \cite{Hilleletal2023}, we presented results for the case in which neither is included. 

\begin{figure*} 
\centering
\includegraphics[trim=0.0cm 1.7cm 0.0cm 0.0cm ,clip, angle=0, scale=0.78]{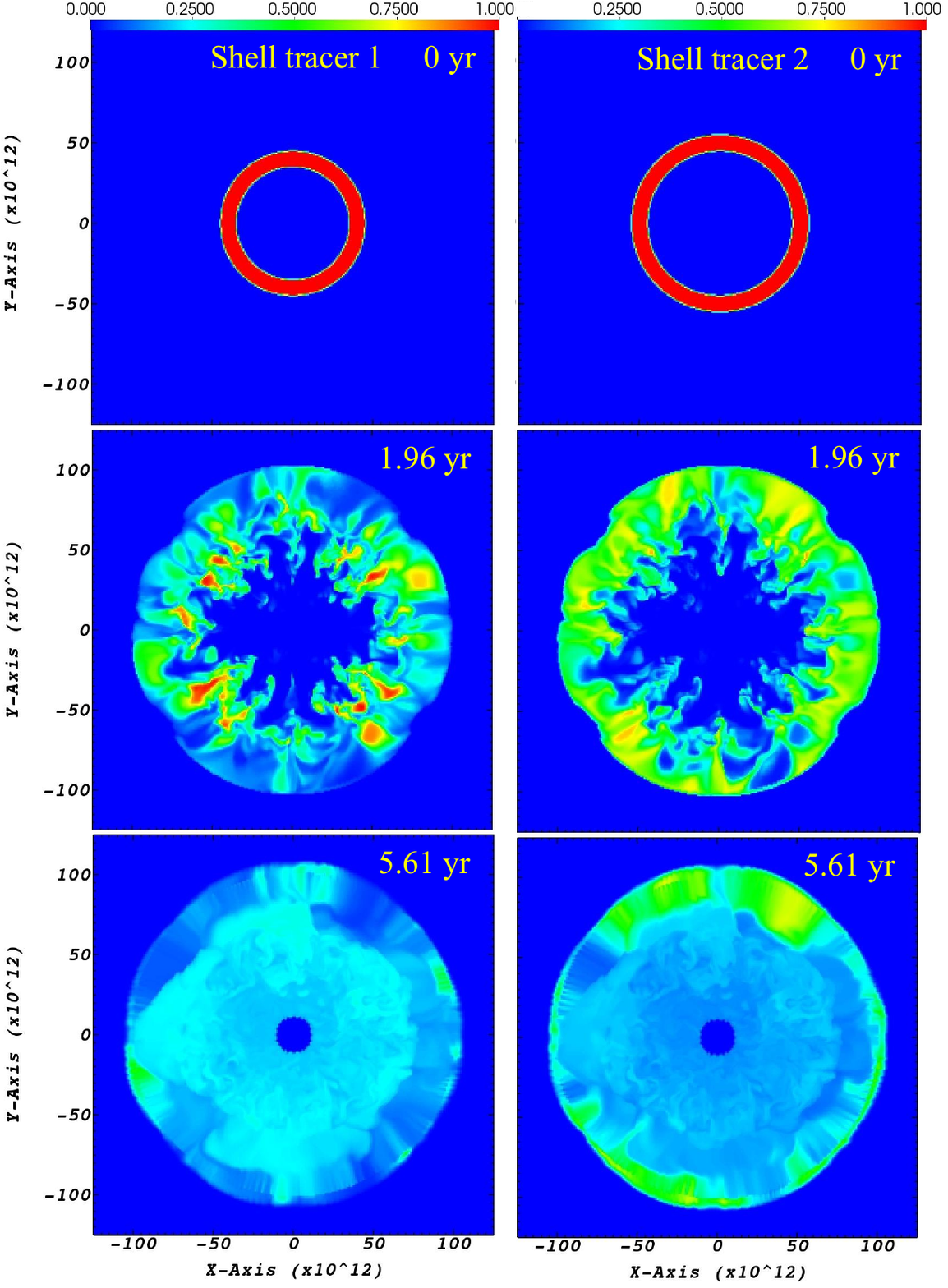}
\caption{Maps of the two tracers in the energy-deposition simulation starting in the inner shell $503 R_\odot < r <  647 R_\odot$ (Shell 1; left column), and in the outer shell $647R_\odot < r <  790 R_\odot$ (Shell 2; right column) in the plane $z=0$, at three times as indicated. We recall that we set a numerical spherical boundary at $R_{\rm w} =100 \times 10^{12} \cm$ (Section \ref{sec:Methods}); this is why the tracer does not expand beyond that radius. 
}
\label{fig:TracerMaps}
\end{figure*}

To present some effects of stellar pulsation and envelope convection, we start by following the average radius of the material starting in two shells. We mark the material in each of the two stellar shells at $t=0$ with a tracer, which is a numerical quantity that follows the flow of a designated material. We designate one tracer to the material starting in the shell $35 \times 10^{12} \cm = 503 R_\odot < r <  45 \times 10^{12} \cm = 647 R_\odot$, and another to the material starting in the shell $45 \times 10^{12} \cm = 647R_\odot < r <  55 \times 10^{12} \cm = 790 R_\odot$.  
Each tracer starts with a value of 1 in the respective shell and zero outside the shell, as we show in the upper panels of Figure \ref{fig:TracerMaps}. Over time, convection mixes the shell material with the rest of the envelope, and the tracer value in a grid cell, between 0 and 1, indicates the fraction of the material that started in the respective shell in that cell. The middle and bottom rows of Figure \ref{fig:TracerMaps} present the tracer maps of the energy-deposition simulation in the plane $z=0$ at two later times: $t= 1.96 \yr$, and $t= 5.61 \yr$. The left column is for the tracer starting in the inner shell, and the right column is for the tracer starting in the outer shell. The tracer maps of the no-energy simulation are very similar in their general structures (differences are in the exact location of clumps).  These panels show the near-complete mixing of the shell within $\simeq 3 \yr$, about four dynamical times $\tau_{\rm d} = (G \bar \rho_0)^{-1/2}= 0.76 \yr$, where $\bar \rho_0$ is the average density of the entire initial stellar model. 
The mixing is a real process and not a numerical artifact. It results from the convection that covers the entire envelope. Later we show that the convection reaches velocities of $\simeq 30-50 \km \s^{-1}$, but in some areas (changing stochastically) the velocity is very low. The stellar radius divided by the average convective velocity gives $881 R_{\odot}/20\km \s^{-1} \simeq 1 \yr$. The convection can account for the mixing in about three years. Because the two tracers mix in about three years, there is no point in adding more tracers. 

\begin{figure} 
\centering
\includegraphics[trim=0.0cm 16.5cm 0.0cm 0.0cm ,clip, angle=0, scale=0.38]{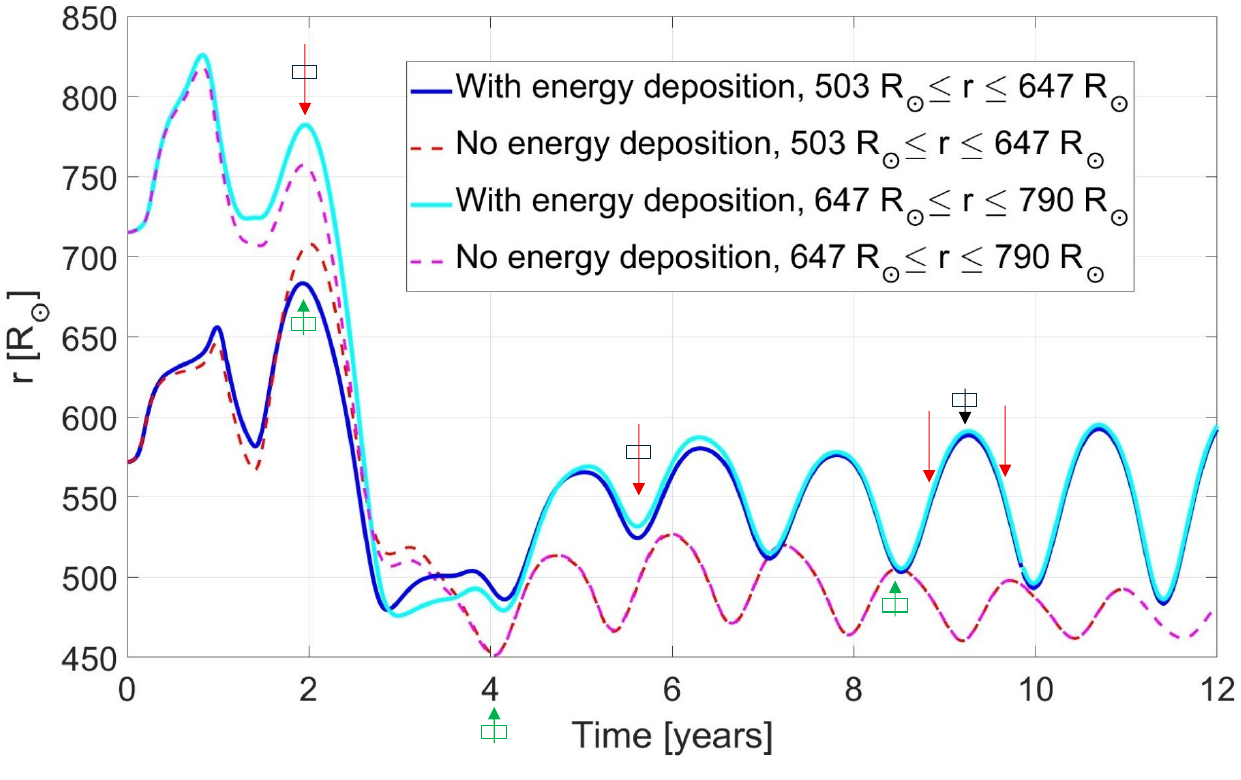}
\caption{The average radii of the tracers of two initial spherical shells: $35 \times 10^{12} \cm = 503 R_\odot < r <  45 \times 10^{12} \cm = 647 R_\odot$ and  $45 \times 10^{12} \cm = 647R_\odot < r <  55 \times 10^{12} \cm = 790 R_\odot$. Dashed lines are for the no-energy simulation, while solid lines are for the simulation with energy deposition. By $t \simeq 3 \yr$, convection mixes the two shells almost completely, and the material that started in separate shells becomes indistinguishable. Without energy deposition, the stellar pulsation decays, while energy deposition in the more realistic simulation maintains the pulsation with a gradual increase in amplitude. The arrows point to different times, at which we later present density and velocity maps.    
}
\label{fig:ShellsRadii}
\end{figure}

The tracer allows us to follow the average radius of the material that started in each shell. 
In Figure \ref{fig:ShellsRadii}, we present the average radius of the material that started in the above-mentioned two shells in the simulation with energy deposition. 

\begin{figure} [t]
\centering
\includegraphics[trim=0.0cm 0.5cm 0.0cm 0.0cm ,clip, angle=0, scale=0.38]{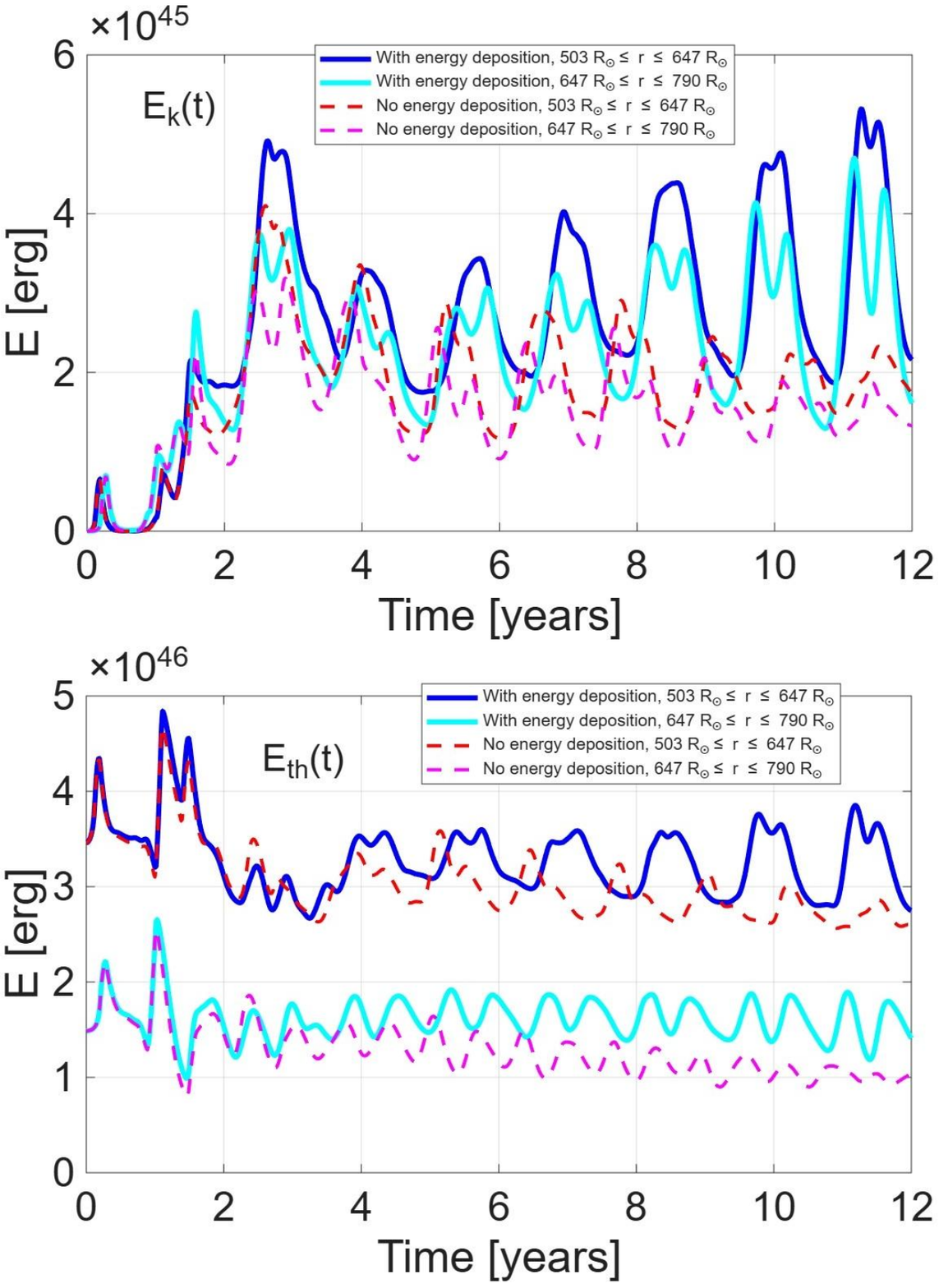}
\caption{The kinetic (upper panel; units of $10^{45} \erg$) and thermal energy (lower panel; units of $10^{46} \erg$) of the two simulations inside two fixed shells: 
$35\times 10^{12} \cm = 503 R_\odot  <r < 45\times 10^{12} \cm = 647 R_\odot$
and 
$45\times 10^{12} \cm = 647 R_\odot  <r < 55\times 10^{12} \cm = 790 R_\odot$. 
The kinetic energy line of each simulation of the inner shell is above that of the outer shell because of the larger mass in the inner shell; the thermal energy line of the inner shell is much higher than that of the outer shell because of the higher temperature of the inner shell.   Solid lines are for the simulation with energy deposition, and dashed lines are for the simulation without energy deposition. This graph further emphasizes stellar pulsation. The pulsation decays without energy deposition. In the more realistic case of energy deposition, stellar pulsation does not decay but rather increases somewhat. The splitting of the peaks, we speculate, suggests pulsation in two or more modes.   
}
\label{fig:OscillationEnergy}
\end{figure}

Figure \ref{fig:ShellsRadii} exhibits several significant properties of our stellar models. 
\begin{enumerate}
    \item The average radius of the material changes its characteristic after two oscillations, with a transition time from $t \simeq 2 \yr$ to $t \simeq 4 \yr$. We note that within the transition time interval, at $t \simeq 3 \yr$, the material in the two shells merges, and the shells become indistinguishable. This is the time period during which convection (turbulence) mixes the envelope. This occurs at about four times the dynamical time of the initial stellar model $\tau_{\rm d} = (G \bar \rho_0)^{-1/2}= 0.76 \yr$. 
    \item The stellar model oscillates (pulsates) with a typical period of $T_{\rm os} \simeq 1 \yr$, which is about the dynamical time of the stellar model. The energy-deposition simulation has a somewhat larger average stellar radius, and hence the dynamical time is longer, and so is the fundamental pulsation mode.    
    \item In \cite{Hilleletal2023}, where we mimicked neither the photospheric emission nor the energy due to nuclear burning in the core, the oscillations decayed after two pulsations, at $t \simeq 3 \yr$, after about the convection mixing time. The cooling of the gas outside the photosphere, mimicking photospheric emission, allows the oscillations to persist for a long time, although they slowly decay as evident from Figure \ref{fig:ShellsRadii}.  
    \item When we include both photospheric cooling and energy deposition, which is the most realistic case we study, the stellar model is larger and the oscillations do not decay; their amplitude even increases somewhat. This is also evident from the evolution of the kinetic energy of the oscillations that we present in Figure \ref{fig:OscillationEnergy}: In the simulation with energy deposition, the amplitudes of the kinetic energy oscillations increase (upper panel solid lines), while without energy deposition, they decrease (upper panel dashed lines). 
\end{enumerate}

In Figure \ref{fig:OscillationEnergy}, we present the kinetic (upper panel) and thermal (lower panel) energies inside two fixed shells as indicated in the panels, for the simulation with (solid lines) and without (dashed lines) energy deposition. This figure shows that, in the more realistic case with energy deposition, the kinetic amplitudes increase somewhat, and the star undergoes significant oscillations, as RSG stars do. Without energy deposition, the oscillation decays.  
In the energy-deposition simulation, the oscillation of the kinetic energy (and, to some extent, the thermal energy) shows two subpeaks at the main peaks and troughs. The average radius of the tracer we present in Figure \ref{fig:ShellsRadii} does not show this behavior. We suspect the splitting of the peaks in the energy graph results from non-radial oscillation modes on top of the main fundamental (radial) mode.  This hypothesis and  the non-radial oscillations are subjects of a future study.   
 
Figure \ref{fig:OscillationEnergy} shows that the kinetic energy is an order of magnitude smaller than the thermal energy. The thermal energy in the energy-deposition simulation oscillates at late times ($t>4 \yr$) around a roughly constant value. It decreases slightly, while the kinetic energy increases somewhat. We checked that the sum of the kinetic and thermal energy oscillates around a constant value at late times. Figure \ref{fig:ShellsRadii} shows that the matter oscillates around a constant radius, indicating conservation of average gravitational energy. Overall, energy is conserved in the energy-deposition simulation. As expected, when we apply cooling without heating, energy decreases (the dashed lines in Figure \ref{fig:OscillationEnergy}).   

We comment on the temperature to which we set the low-density cells. We cooled the cells with density below the initial photospheric density to $T_{\rm ch}= 1000 \K$. Since the initial photospheric temperature (1D model) is only $\simeq 3150 \K$, imposing substantial cooling requires a low temperature. Hence, we did not use a cooling temperature higher than $T_{\rm ch}= 1000 \K$. We already know that without cooling, the oscillations rapidly decay \citep{Hilleletal2023}. When we inject energy to mimic the nuclear burning in the core, we must remove energy from the surface. Therefore, we cannot inject energy without substantial cooling. We did not use a cooling temperature below $T_{\rm ch}= 1000 \K$ to avoid overly low pressures in some cells that might lead to unphysical flow structures. We will study the influence of this temperature on the results in the future, but it will require several very long simulations.

\begin{figure*} 
\centering
\includegraphics[trim=0.0cm 6.5cm 0.0cm 2.2cm ,clip, angle=0, scale=0.80]{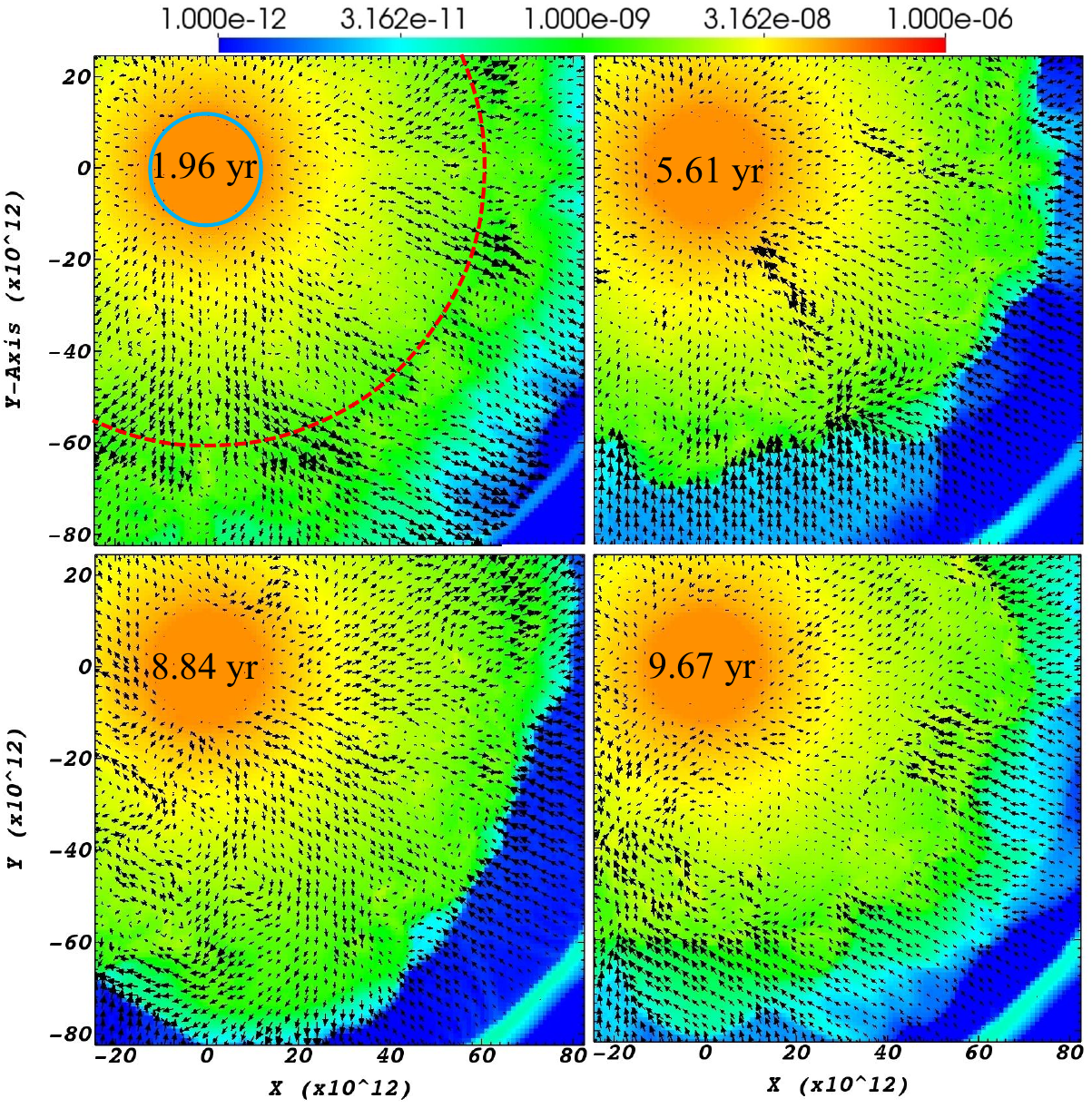}
\caption{Velocity and density maps in the plane $z=0$ for the energy-deposition simulation at four times as indicated. The four times of these panels are marked with red down-pointing arrows in Figure \ref{fig:ShellsRadii}. The pale-blue circle is the inert core, and the red-dashed line is the initial surface of the giant. The initial stellar radius is five times the inert core: $R_{\ast,0}=881 R_\odot = 61 \times 10^{12} \cm$. The arrows depict the velocities: direction and magnitude relative to the arrow length. The maximum velocity value at each time is $55 \km \s^{-1}$,  $47 \km \s^{-1}$, $71 \km \s^{-1}$, $45 \km \s^{-1}$, respectively. The colors depict the density according to the color bar in units of $\g \cm^{-3}$.  These panels highlight the efficient convective motion and pulsation. The pale-blue arc on the lower right of the panels is due to the spherical numerical boundary at $R_{\rm w} =100 \times 10^{12} \cm$. 
}
\label{fig:VelocityMaps}
\end{figure*}

\begin{figure*} 
\centering
\includegraphics[trim=0.0cm 0.0cm 0.0cm 0.0cm ,clip, angle=0, scale=0.65]{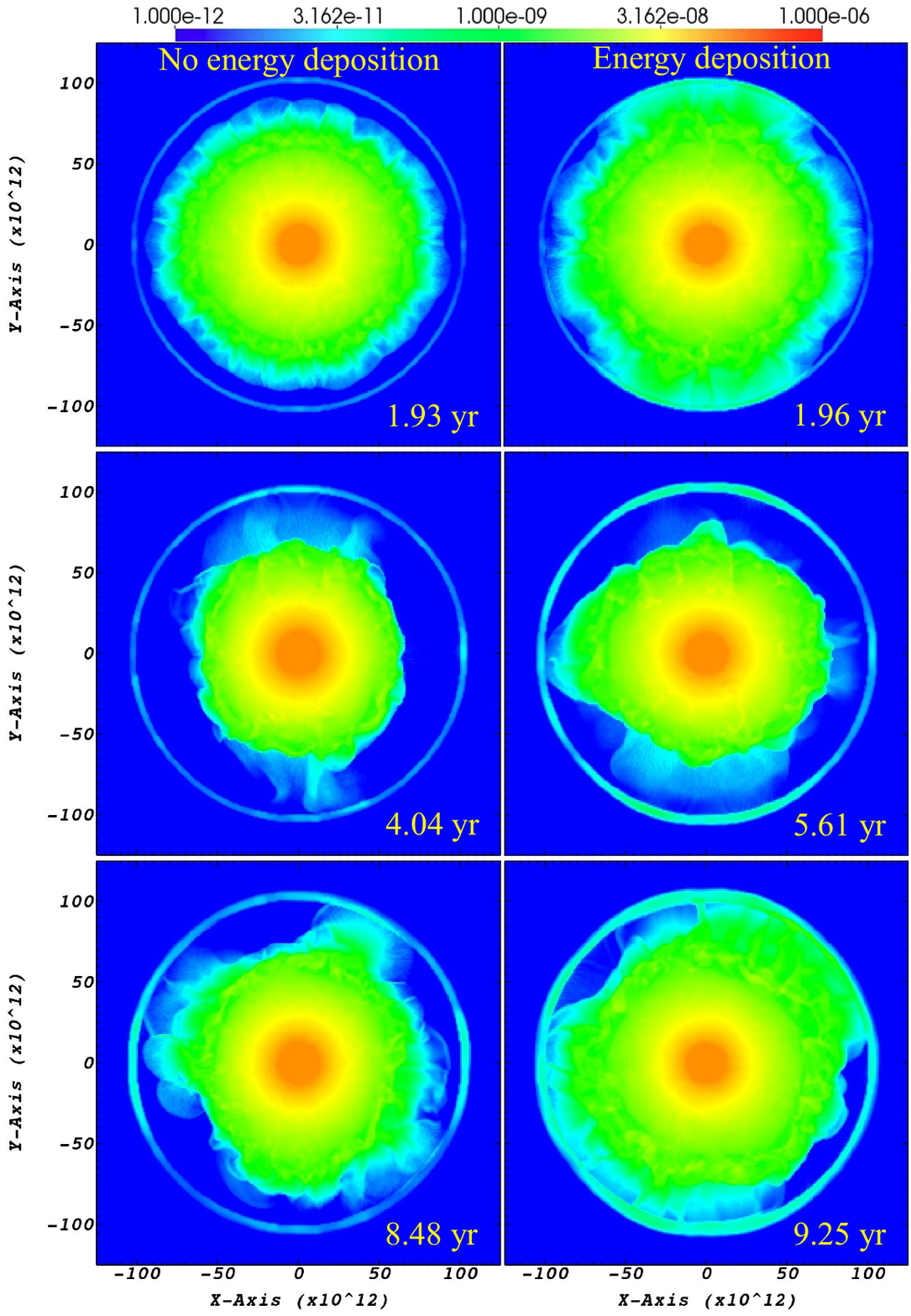}
\caption{Density maps in the plane $z=0$ for the No-energy-deposition simulation (left column) and the energy deposition simulation (right column), each at three times; note that times differ between the two simulations. We mark the times of these six panels in Figure \ref{fig:ShellsRadii}: for the no-energy simulation with the green up-pointing arrows with rectangles, and for the energy simulation with down-pointing arrows with black rectangles. 
The colors depict the density according to the color bar in units of $\g \cm^{-3}$. These panels highlight, what we speculate are, non-spherical oscillations of the stellar model, with a larger amplitude in the energy deposition simulation. The pale-blue shell (seen as a ring) outside the stellar model is a numerical artifact; it has a very low density and a negligible dynamical effect. The pale-blue circle is due to the spherical numerical boundary at $R_{\rm w} =100 \times 10^{12} \cm$. It has a very low density, well below the photospheric density.   
}
\label{fig:DensityMaps}
\end{figure*}

In Figure \ref{fig:VelocityMaps}, we present the velocity and density maps in the plane $z=0$ for the energy deposition simulation. The four panels are at the times marked by red down-pointing arrows in Figure \ref{fig:ShellsRadii}. The longest arrows in the panels depict velocities of $\simeq 30-50 \km \s^{-1}$, or the order of the Keplerian velocity on the surface of the initial model ($v_{\rm Kep}= 52 \km \s^{-1}$). This indicates the development of vigorous convection, as expected. 
The first panel ($t=1.96 \yr$) is before full convective mixing and during stellar expansion. Most arrows point outwards, but at different magnitudes, and many arrows point in other directions, indicating the presence of convection. The three panels at later times show changes in stellar shape and radius (i.e., pulsation) and the persistence and erratic nature of convection. The stellar outer boundary (the green-blue boundary) is not circular, indicating that there are modes beyond the fundamental radial pulsation mode, i.e.,  hinting at non-radial pulsational modes. We further present the structures resulting from what we speculate are non-radial pulsational modes in Figure \ref{fig:DensityMaps}, for the no-energy simulation in the left column and the energy-deposition simulation in the right column. The first and last rows are at maximum expansion, and the middle row is at minimum, as the respective arrows with rectangles show in Figure \ref{fig:ShellsRadii}. The non-radial pulsational modes are the subject of a future study and may require higher grid resolution.  

The density maps present a thin, low-density shell outside the stellar model. As we explained in Section \ref{sec:Methods}, we set a spherical numerical boundary at $R_{\rm w}=100 \times 10^{12} \cm = 1437 R_\odot= 1.63 R_{\ast,0}$ by imposing a very slow outward velocity there of $0.1 \km \s^{-1}$. The velocity maps (Figure \ref{fig:VelocityMaps}) show that the velocity is very low from that radius outward. 
The study of the best boundary condition that will also mimic stellar wind is a subject of a future paper. 

\section{Higher resolution simulation and convergence}
\label{sec:Convergence}

We examined convergence by performing the energy-deposition simulation at a higher resolution. 
Instead of a cell size of $\Delta _{\rm c}=L_{\rm G}/256 = 9.766 \times 10^{11} \cm$ and a grid size of $L_{\rm G} = 250 \times 10^{12} \cm$ in the simulations we described above, we performed the high-resolution energy-deposition simulation with $\Delta _{\rm c}=L_{\rm G}/512 = 4.8828 \times 10^{11} \cm$ and the same grid size.
Because of the higher resolution, the simulation proceeds extremely slowly, and we could not continue it beyond about four years of simulated time.

\begin{figure} 
\centering
\includegraphics[trim=0.0cm 19.5cm 0.0cm 0.0cm ,clip, angle=0, scale=0.50]{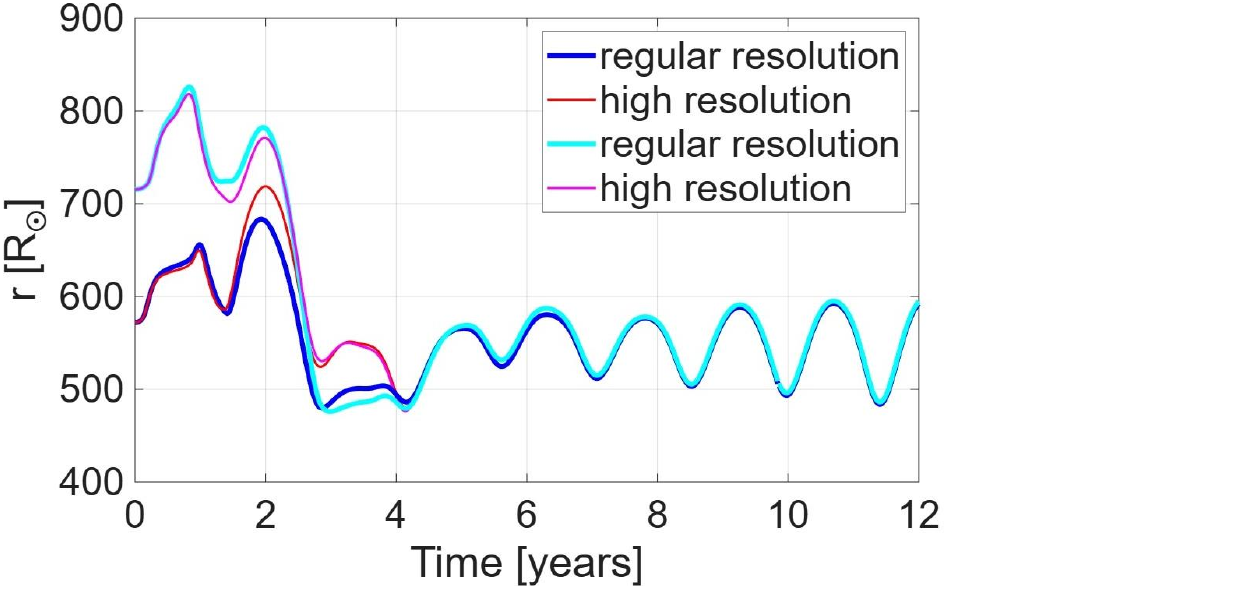}
\caption{ The average radii of the tracers of two initial spherical shells: $35 \times 10^{12} \cm = 503 R_\odot < r <  45 \times 10^{12} \cm = 647 R_\odot$ and  $45 \times 10^{12} \cm = 647R_\odot < r <  55 \times 10^{12} \cm = 790 R_\odot$, for the energy deposition simulations: thick lines: regular resolution as in Figure \ref{fig:ShellsRadii}; thin lines: the high resolution simulation. 
}
\label{fig:ShellsRadiiResolution}
\end{figure}

In Figure \ref{fig:ShellsRadiiResolution}, we present the average radii of the material that started in the two shells, as in Figure \ref{fig:ShellsRadii}, for the simulation with energy deposition at regular resolution (thick lines) and the high-resolution simulation (thin lines). The results are very similar up to $t =2.75 \yr$, then diverge somewhat, and then merge again. As we discussed in our description of Figure \ref{fig:ShellsRadii}, the average radii of the material in the regular-resolution simulation have a transition time from $t \simeq 2 \yr$ to $t\simeq 4 \yr$; at $t\simeq 4 \yr$, the shells start regular oscillations. From the high-resolution simulation, it seems that the shells start regular oscillations early on, at $t\simeq 3 \yr$. 

\begin{figure} [t]
\centering
\includegraphics[trim=0.0cm 9.5cm 0.0cm 0.0cm ,clip, angle=0, scale=0.50]{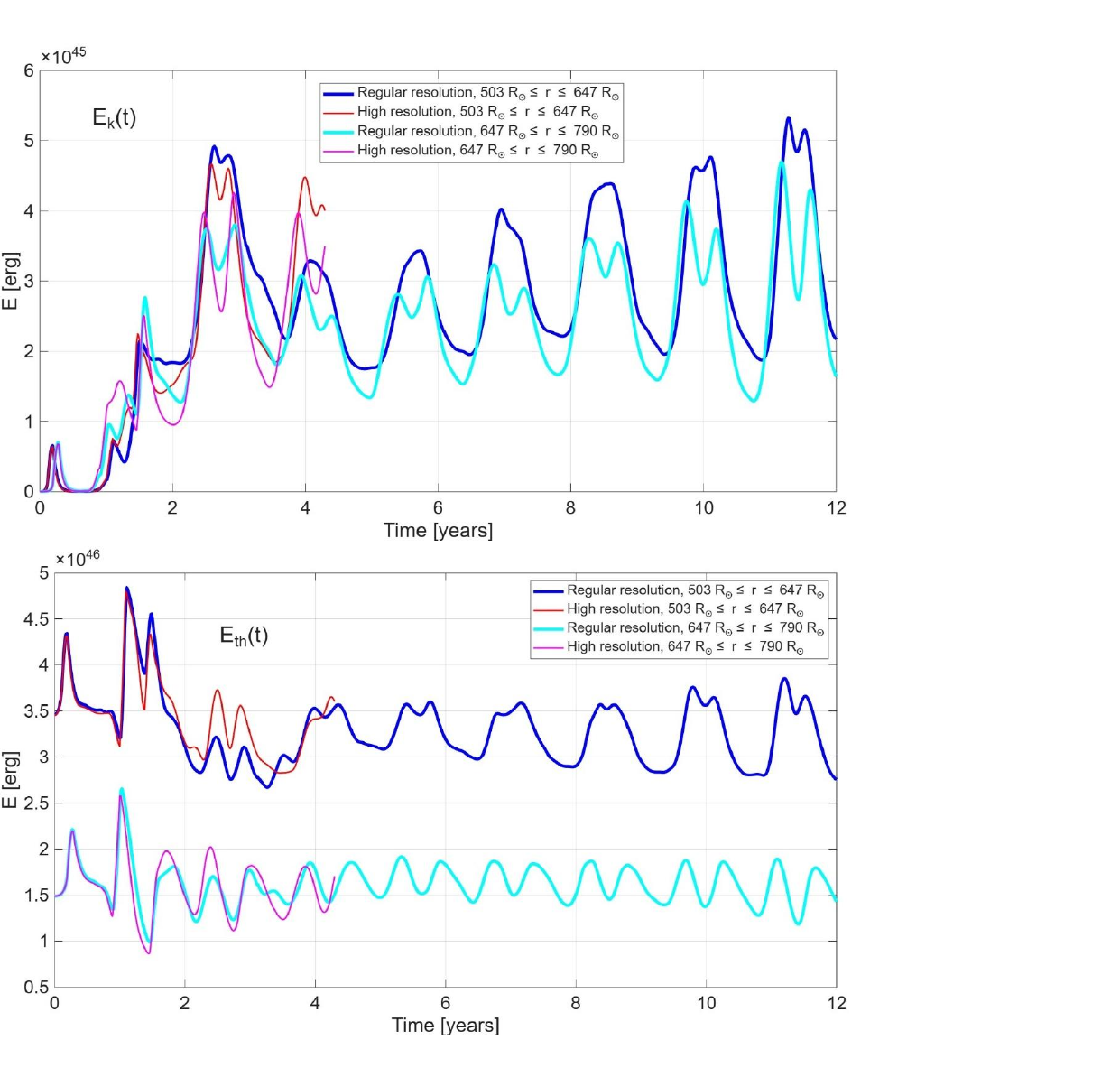}
\caption{ Comparing the kinetic energy evolution (upper panel) and thermal energy evolution (lower panel) of the regular-resolution (thick lines) and high-resolution (thin lines) simulations with energy deposition (similar to Figure \ref{fig:OscillationEnergy}). The upper lines are for material in the shell $503 R_\odot < r < 647 R_\odot$, and the lower lines are for material in the shell $647 R_\odot < r < 790 R_\odot$.
}
\label{fig:OscillationEnergyResolution}
\end{figure}

In Figure \ref{fig:OscillationEnergyResolution}, we present the kinetic and thermal energy inside two shells, similar to Figure \ref{fig:OscillationEnergy}. The thin lines of the high-resolution simulation with energy deposition are close to those of the regular simulation (thick lines); namely, the results are similar. The kinetic energy lines of the high-resolution simulations are above those of the normal resolution around $t=4 \yr$, but are at the same height as the regular resolution at later times. This again shows that the high-resolution simulations start exhibiting the regular oscillation earlier. Overall, the regular resolution seems to have reached convergence to an accuracy that satisfies us.
   
The similarity of the oscillatory behavior in the two resolutions suggests that the oscillations are physical oscillations. The initial perturbations arise from importing the 1D model onto the 3D grid. However, the oscillations are maintained by the energy deposition scheme, as evident from the increase in the amplitude of the kinetic energy in Figure \ref{fig:OscillationEnergy} (solid lines), and from previous simulations that show that without energy deposition the oscillations rapidly decay \citep{Hilleletal2023}.  

In addition, we checked that the density and pressure profiles (averaged over spherical shells and during the oscillations phase after several years) of the 3D simulation with regular resolution are very similar to the 1D model that we transported to our 3D numerical grid; also showing that the regular resolution is adequate for our goals. 

\begin{figure} [t]
\centering
\includegraphics[trim=0.0cm 0.0cm 0.0cm 0.0cm ,clip, angle=0, scale=0.50]{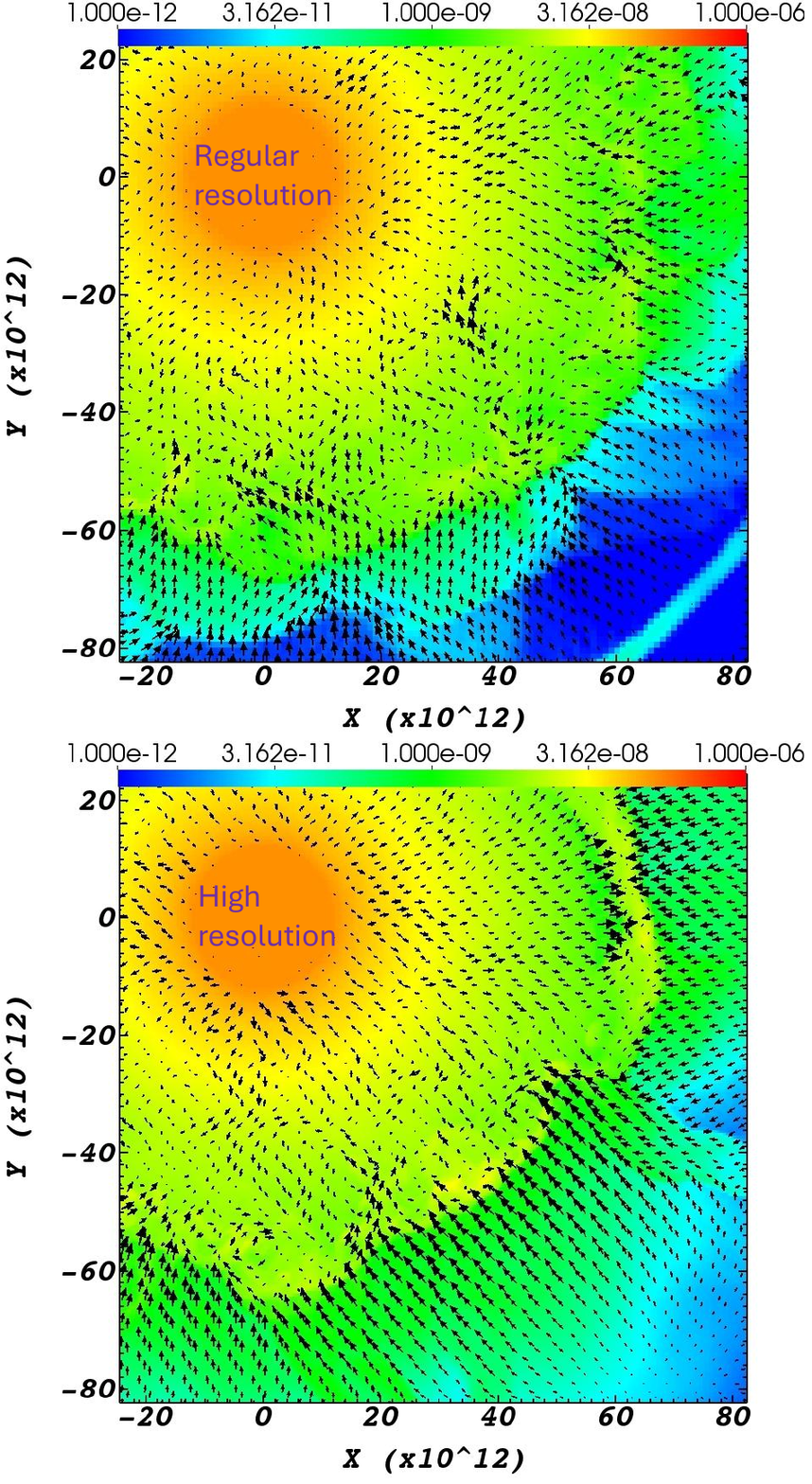}
\caption{ Velocity and density maps in the plane $z=0$, similar to Figure \ref{fig:VelocityMaps}, but for the regular resolution (upper panel) and high resolution (lower panel), both with energy deposition and at $t=3.8 \yr$. The maximum velocity in the regular simulation map (upper panel) is $37 \km \s^{-1}$, and in the high-resolution simulation $50 \km \s^{-1}$.   
}
\label{fig:HighResVmaps}
\end{figure}

Finally, in Figure \ref{fig:HighResVmaps} we present the velocity and density maps of the regular-resolution and high-resolution simulations with energy deposition, both at $t=3.8 \yr$.  The images are not identical, but sufficiently similar. We cannot expect identical maps because the vorticity is stochastic and is expected to change from one grid to another. It is not a matter of convergence, but of stochastic behavior, whose evolution depends on initial perturbations, which in turn depend on the numerical grid. The convection affects the exact oscillation behavior, but not the general behavior, as evident from Figures  \ref{fig:ShellsRadiiResolution} and \ref{fig:OscillationEnergyResolution}

\section{Summary}
\label{sec:Summary}

The convection and pulsation of RGB, AGB, and RSG stars influence the CEE in several ways. 
The pulsation can lead to early, intermittent interaction, in which the envelope covers the companion on and off before a full CEE begins. Alternatively, this intermittent engulfment can facilitate the grazing envelope evolution. The pulsation also plays a role when the companion is deep inside the envelope, as the gravity of the companion can excite some oscillation modes, even by a sub-stellar companion (e.g., \citealt{Soker1993}). Convection can change the accretion process onto the companion star, e.g., by introducing a stochastic component to the angular momentum of the material that the companion accretes. This, in turn, can lead to wobbling jets if the companion launches jets (e.g., \citealt{Dorietal2023, Hilleletal2026}). For these reasons, it is important to build a 3D stellar model as realistic as possible for CEE simulations. In earlier studies, we did not relax the stellar model we imported from a 1D simulation and found the development of convection and pulsation. However, in that model, the pulsation decayed after two oscillations, about three years. 

In this study, we mimic the photospheric radiation by cooling to $1000 \K$ any numerical cell with a density below the initial photospheric density, and we mimic the nuclear energy production by depositing energy in an inner shell above the inert core at the same power as the 1D stellar model luminosity. These two very simple ingredients reproduce much more realistic pulsations. However, we did not study the sensitivity of our results to the cooling temperature; future studies of our method should examine temperatures below and somewhat above $1000 \K$.      

We found that mimicking stellar radiation alone prolongs the decay time of the pulsations much beyond two pulsations, but they slowly decay (dashed lines in Figure \ref{fig:ShellsRadii} for the average radius of the material and in Figure \ref{fig:OscillationEnergy} for the kinetic and thermal energy in two shells). Adding energy deposition to mimic stellar luminosity maintains these pulsations, and their amplitude slowly grows. 

The splitting of the peaks and troughs, mainly in the kinetic energy lines of the simulation with energy (solid lines in the upper panel of Figure \ref{fig:OscillationEnergy}), and the non-spherical surface of the stellar model (Figures \ref{fig:VelocityMaps} and \ref{fig:DensityMaps}) hint  on the presence of non-radial oscillation modes on top of the fundamental radial mode. The study of pulsation modes is a subject for future study that may require higher-resolution simulations. 

The tracer maps for two shells in the simulation with energy deposition show how convection mixes the envelope gas. The average radii of material in these shells (Figure \ref{fig:ShellsRadii}) show that the mixing time is about 3 years, about four dynamical times. A more extended quantitative study of the convection is the subject of a forthcoming paper. 

Figures \ref{fig:ShellsRadiiResolution} - \ref{fig:HighResVmaps}, which compare the regular resolution with a high-resolution simulation for about four years, show that the regular simulations we have used reached convergence to the accuracy that satisfies our demands.

Future studies should examine the influence of the inner boundary at the inert core's outer surface and the effect of the inert core's size on pulsation modes (we expect that a smaller core will allow some more higher-harmonic pulsation modes). We cannot decrease this radius by much due to numerical limitations.  
Future studies should also explore the best outer boundary conditions near the stellar surface to mimic the stellar wind, as these giant stars have very intense winds, particularly if interacting with a close stellar companion. 

Our conclusion, which still needs further studies, i.e., with higher resolutions, is that we can prepare a giant star to simulate CEE and grazing-envelope evolution by depositing energy with the stellar luminosity in an inner shell, 
and cooling the outer low-density numerical shell. In this approach, we do not need to relax the model. The pulsations and vigorous convection that our approach obtains provide a better fit to simulations of the onset of the CEE and grazing-envelope evolution.  

\section*{Acknowledgements}
We thank an anonymous referee for detailed and thorough comments and suggestions that substantially improved the manuscript. 
A grant from the Pazy Foundation (2026) supported this research.

 \bibliography{bib}{}
  \bibliographystyle{aasjournal}

\end{document}